# Implications of Form Invariance to the Structure of Nonextensive Entropies

A. K. Rajagopal[1] and Sumiyoshi Abe[2]

[1]*Naval Research Laboratory, Washington, DC 20375-5320, USA*

[2]*College of Science and Technology, Nihon University,*

*Funabashi, Chiba 274-8501, Japan*

The form invariance of the statement of the maximum entropy principle and the metric structure in quantum density matrix theory, when generalized to nonextensive situations, is shown here to determine the structure of the nonextensive entropies. This limits the range of the nonextensivity parameter $q$ to $(0,1)$ so as to preserve the concavity of the entropies. The Tsallis entropy is thereby found to be appropriately renormalized.

PACS numbers: 05.20.-y, 05.30.-d

Nonextensive generalization of Boltzmann-Gibbs statistical mechanics is found to be essential for a cogent description of a large number of observed phenomena involving (multi)fractal structures, long-range interactions, and long-time memories, which are increasingly of general interest in many facets of physical sciences. In particular, the form of the generalization originally proposed by Tsallis [1] has been widely used to analyze many such systems, phenomena, and theories. Examples are Lévy-type random walks and anomalous diffusion [2], hydrogen-atom specific heat [3], pure-electron plasma [4], stellar polytropes [5], mean-field magnetization [6], velocity distribution of clusters of galaxies [7], interpolative quantum statistics [8], and generalized theory of thermal Green's functions [9]. (A comprehensive list of references to the literature on this subject is currently available at URL [10].) However, some ambiguities in the early proposal have remained and attempts have been made to rectify them by introducing appropriate definitions of generalized expectation values in order to be in conformity with the conventional properties of statistical principles [11]. These modifications have been more or less *ad hoc*, thus inviting investigations of guiding principles which would lead to a clear basis for any proposed forms for the generalizations of Boltzmann-Gibbs statistical mechanics.

In this paper, it is suggested that the idea of the relative entropy as propounded by Kullback and Leibler [12] can provide such a guiding principle for determining unambiguously the structure of the nonextensive entropies. This is because the Kullback-Leibler (KL) entropy gives a measure of uncertainty in the objective state $\hat{\rho}$ with respect to the reference state $\hat{\rho}'$:

$$K_1[\hat{\rho}, \hat{\rho}'] = \mathrm{Tr}\left[\hat{\rho}\left(\ln \hat{\rho}'^{-1} - \ln \hat{\rho}^{-1}\right)\right] \geq 0, \qquad (1)$$

where $\hat{\rho}$ and $\hat{\rho}'$ are normalized density matrices. The equality holds if and only if $\hat{\rho} = \hat{\rho}'$. Therefore $K_1[\hat{\rho}, \hat{\rho}']$ quantifies the difference between $\hat{\rho}$ and $\hat{\rho}'$. It is a physically measurable quantity unlike the entropy itself and serves to affirm the nature of maximum entropy principle [13] as well as an underlying geometric structure [14]. We will discuss here a similarly defined relative entropy for the nonextensive systems along with the appropriate definition of the normalized $q$-expectation values. The central result of this analysis is that the original Tsallis entropy should be properly modified in order to establish the form invariance by presenting the generalizations of the above two properties. The KL entropy in eq. (1) and the von Neumann entropy in eq. (2) below are presented in these particular ways in anticipation of the form invariance that we are seeking when they are generalized. Unlike the original Tsallis entropy, this modified form is found to be concave only if the nonextensivity parameter $q$ lies in the interval $(0,1)$, into which all of the nonextensive Hamiltonian systems seem to fall [15].

We begin by deriving the above-mentioned two properties in the Boltzmann-Gibbs context so as to provide us with the guiding equations, with which the results of our generalization should conform. We do this in the quantum-mechanical context by maximizing the von Neumann entropy

$$S_1[\hat{\rho}] = \text{Tr}(\hat{\rho} \ln \hat{\rho}^{-1}) \qquad (2)$$

subject to the constraints of the expectation value of the energy $U_1 = \text{Tr}(\hat{\rho}\hat{H}) \equiv \langle \hat{H} \rangle_1$ (where $\hat{H}$ is the Hamiltonian operator of the system under consideration) and the normalization of the density matrix. Throughout this paper, we use units with the Boltzmann constant $k_B$ set equal to unity. The resulting density matrix is found to be $\hat{\rho}' = \exp[-\beta(\hat{H} - U_1)]/Z_1(\beta)$. Here, $\beta$ is the Lagrange multiplier associated with the energy constraint which is identified with the inverse temperature, and $Z_1(\beta) = \text{Tr}\exp[-\beta(\hat{H} - U_1)]$ is equivalent to the standard partition function. Using this in eq. (1), we have the following maximum entropy relation:

$$K_1[\hat{\rho}, \hat{\rho}'] = S_1[\hat{\rho}'] - S_1[\hat{\rho}] + \beta \langle \hat{H} - U_1 \rangle_1 \geq 0. \qquad (3)$$

This is the first of the equations, the form of which we wish to preserve in our nonextensive generalization.

On the other hand, suppose the density matrices be characterized by a parameter $\alpha$. (Extension to the case of multiparameter is straightforward.) Since $\hat{\rho}(\alpha)$ is a Hermitian, positive semi-definite, traceclass operator, it can be expanded in terms of its orthonormal eigenbasis $\{|a(\alpha)\rangle\}$ as follows:

$$\hat{\rho}(\alpha) = \sum_a p_a(\alpha)|a(\alpha)\rangle\langle a(\alpha)|. \qquad (4)$$

The expansion coefficients are usually interpreted as the probabilities of finding the system in the eigenbasis states $\{|a(\alpha)\rangle\}$ and therefore satisfy the relations

$$0 \leq p_a(\alpha) \leq 1, \quad \sum_a p_a(\alpha) = 1. \qquad (5)$$

Taking $\hat{\rho}' = \hat{\rho}(\alpha)$ and $\hat{\rho} = \hat{\rho}(\alpha + d\alpha)$ (where $d\alpha$ is an infinitesimal change of $\alpha$) in eq. (1), up to the second order in $d\alpha$, we find the following metric:

$$ds^2 = K_1[\hat{\rho}, \hat{\rho}'] + K_1[\hat{\rho}', \hat{\rho}] = ds_{cl}^2 + ds_{qu}^2, \qquad (6)$$

$$ds_{cl}^2 = (d\alpha)^2 \sum_a \frac{[\partial_\alpha p_a(\alpha)]^2}{p_a(\alpha)} = 4(d\alpha)^2 \langle\langle [\partial_\alpha \ln p_a^{-1/2}(\alpha)]^2 \rangle\rangle_1, \qquad (7)$$

$$ds_{qu}^2 = 2(d\alpha)^2 \sum_{a,a'} \left|\langle a'(\alpha)|\overleftrightarrow{\partial}_\alpha|a(\alpha)\rangle\right|^2 p_a(\alpha)\left[\ln p_{a'}^{-1}(\alpha) - \ln p_a^{-1}(\alpha)\right]$$

$$= 2(d\alpha)^2 \left\langle\!\!\left\langle \sum_{a'} \left|\langle a'(\alpha)|\overleftrightarrow{\partial}_\alpha|a(\alpha)\rangle\right|^2 \left[\ln p_{a'}^{-1}(\alpha) - \ln p_a^{-1}(\alpha)\right]\right\rangle\!\!\right\rangle_1, \quad (8)$$

where the double bracket symbol stands for the ordinary expectation value with respect to $p_a(\alpha)$. This is the second of the equations, the form of which we wish to preserve in our nonextensive generalization. We note that equation (7) does not contain the overlap between basis vectors, resembling the classical Fisher metric (i.e., the KL divergence) even though it has quantum probabilities, whereas equation (8) is explicitly quantum in nature by virtue of the presence of the overlap. The latter quantity is seen to be positive, implying the quantum contribution expands the scale of length as it should do. To the best of our knowledge, this observation has not been previously made in the literature.

The structures exhibited in eqs. (3) and (6) will be the two chosen forms which any generalization of $S_1[\hat{\rho}]$ to a nonextensive entropy is required to satisfy. Thus our proposal of a guiding principle is to maintain the "form invariance" of these equations in such a generalization.

The following form of the generalized KL or $q$-KL entropy

$$K_q^{(a)}[\hat{\rho}, \hat{\rho}'] = \mathrm{Tr}\!\left[\hat{\rho}^q\!\left(\mathrm{Ln}_q\,\hat{\rho}'^{-1} - \mathrm{Ln}_q\,\hat{\rho}^{-1}\right)\right]\!\Big/\mathrm{Tr}(\hat{\rho}^q) \geq 0 \quad (9)$$

and the modified form of the Tsallis entropy

$$S_q^{(a)}[\hat{\rho}] = \mathrm{Tr}\!\left(\hat{\rho}^q \mathrm{Ln}_q\,\hat{\rho}^{-1}\right)\!\Big/\mathrm{Tr}(\hat{\rho}^q), \quad (10)$$

where

$$\mathrm{Ln}_q x \equiv \left(x^{q-1} - 1\right)/(q-1), \quad (11)$$

will be shown to lead to form invariant structures in the sense defined above. In the limit $q \to 1-$, equations (9), (10), and the definition of $\mathrm{Ln}_q x$ become eqs. (1), (2), and the standard logarithm, $\ln x$, respectively. Expressions (9) and (10) correspond to those given in Refs. [16,17] and [1] divided by $\mathrm{Tr}(\hat{\rho}^q) \equiv c_q(\hat{\rho})$, respectively. Although the inequality in eq. (9) holds for any positive $q$, the entropy (10) is seen to be concave for $q$ only in the interval $(0, 1)$. The modified expressions (9) and (10) satisfy the $H$-theorem as the original ones do [17]. However, the additivity property of $S_q^{(a)}[\hat{\rho}]$ is

changed to

$$S_q^{(a)}[\hat{\rho}_A \otimes \hat{\rho}_B] = S_q^{(a)}[\hat{\rho}_A] + S_q^{(a)}[\hat{\rho}_B] + (q-1)S_q^{(a)}[\hat{\rho}_A]S_q^{(a)}[\hat{\rho}_B]. \qquad (12)$$

It is of interest to note that this relation has the same structure of the Jackson basic number of $A$ defined by $[A]_q \equiv (q^A - 1)/(q-1)$ in $q$-deformation theory. The basic number of $A+B$ satisfies the identity: $[A+B]_q = [A]_q + [B]_q + (q-1)[A]_q[B]_q$, which has a striking similarity to eq. (12). To understand this similarity better, we consider a quantity $f(x) = 1/\text{Tr}(\hat{\rho}^x)$. Clearly, $f(1) = 1$. The von Neumann entropy is expressed as the rate of infinitesimal translation of the index $x$ from $x=1$: $S_1[\hat{\rho}] = df(x)/dx\big|_{x=1}$. On the other hand, the modified Tsallis entropy is found to be given by the Jackson $q$-differential: $S_q^{(a)}[\hat{\rho}] = D_q f(x)\big|_{x=1} \equiv [f(qx)-f(x)]/(qx-x)\big|_{x=1}$. Convergence of $S_q^{(a)}[\hat{\rho}]$ on $S_1[\hat{\rho}]$ in the limit $q \to 1$ is due to the fact that the Jackson $q$-differential becomes the ordinary differential in such a limit. Equation (11) is a result arising from the $q$-deformed Leibniz rule which the Jackson differential satisfies: $D_q[f(x)g(x)] = [D_q f(x)]g(qx) + f(x)[D_q g(x)]$. These properties are similar to those noted in the case of the original Tsallis entropy [18]. A physical interpretation of this observation is that the system entropy has the scale invariance with respect to the nonextensive parameter $q$.

We choose $\hat{\rho}'$ to be the one which maximizes the $S_q^{(a)}$ subject to the constraints of the normalized $q$-expectation value of the energy $U_q = \text{Tr}(\hat{\rho}^q \hat{H})/c_q(\hat{\rho}) \equiv \langle \hat{H} \rangle_q$ and the normalization of the density matrix. The resulting density matrix is found to be

$$\hat{\rho}' = \frac{1}{Z_q(\beta)}\left[1 - (1-q)\beta c_q\left(\hat{H} - U_q\right)\right]^{1/(1-q)}, \qquad (13)$$

$$Z_q(\beta) = \text{Tr}\left[1 - (1-q)\beta c_q\left(\hat{H} - U_q\right)\right]^{1/(1-q)}, \qquad (14)$$

where the Lagrange multiplier $\beta$ associated with the constraint on the $q$-expectation value of the energy defined above is identified with the inverse temperature. The generalized partition function $Z_q(\beta)$ and the factor $c_q$ satisfy the identity: $c_q = [Z_q(\beta)]^{1-q}$. Using eq. (13) in eq. (9), we have the following maximum entropy relation:

$$K_q^{(a)}[\hat{\rho}, \hat{\rho}'] = S_q^{(a)}[\hat{\rho}'] - S_q^{(a)}[\hat{\rho}] + \beta\langle \hat{H} - U_q \rangle_q \geq 0. \qquad (15)$$

Thus, we see that equation (15) maintains the same form as eq. (3).

Now, taking $\hat{\rho}' = \hat{\rho}(\alpha)$ and $\hat{\rho} = \hat{\rho}(\alpha + d\alpha)$ in eq. (9) and carrying out the calculation as before up to the second order in $d\alpha$, we find the following generalized metric:

$$ds_q^2 = K_q^{(a)}[\hat{\rho}, \hat{\rho}'] + K_q^{(a)}[\hat{\rho}', \hat{\rho}] = ds_{q,cl}^2 + ds_{q,qu}^2, \qquad (16)$$

$$ds_{q,cl}^2 = \frac{q(d\alpha)^2}{\sum_a p_a^q(\alpha)} \sum_a \frac{[\partial_\alpha p_a(\alpha)]^2}{p_a(\alpha)} = 4q(d\alpha)^2 \left\langle\!\!\left\langle \left[\partial_\alpha \mathrm{Ln}_q\, p_a^{-1/2}(\alpha)\right]^2 \right\rangle\!\!\right\rangle_q, \qquad (17)$$

$$\begin{aligned} ds_{q,qu}^2 &= \frac{2(d\alpha)^2}{\sum_a p_a^q(\alpha)} \sum_{a,a'} \left|\langle a'(\alpha)|\overset{\leftrightarrow}{\partial}_\alpha|a(\alpha)\rangle\right|^2 p_a^q(\alpha)\left[\mathrm{Ln}_q\, p_{a'}^{-1}(\alpha) - \mathrm{Ln}_q\, p_a^{-1}(\alpha)\right] \\ &= 2(d\alpha)^2 \left\langle\!\!\left\langle \sum_{a'} \left|\langle a'(\alpha)|\overset{\leftrightarrow}{\partial}_\alpha|a(\alpha)\rangle\right|^2 \left[\mathrm{Ln}_q\, p_{a'}^{-1}(\alpha) - \mathrm{Ln}_q\, p_a^{-1}(\alpha)\right] \right\rangle\!\!\right\rangle_q, \end{aligned} \qquad (18)$$

where the double bracket is now defined by $\langle\!\langle A_a(p_a)\rangle\!\rangle_q \equiv \sum_a p_a^q A_a(p_a) / \sum_a p_a^q$. Thus, we again see that equation (16) maintains the same form as eq. (6).

We have therefore established the two form invariances of the generalization promised in the beginning of the paper.

In conclusion, the form invariance requirement has led us to the modifications of the $q$-KL entropy and the original Tsallis entropy as in eqs. (9) and (10). Had we used the original forms of these quantities (i.e., without division by $c_q$), the form invariances would not be established.

A. K. R. acknowledges the hospitality at Nihon University–Funabashi and Institute for Molecular Science–Okazaki, Japan. He also acknowledges the partial support of the US Office of Naval Research. S. A. was supported in part by School of Liberal Arts, College of Science and Technology, Nihon University.